\documentclass[prl,amsmath,amssymb,twocolumn,superscriptaddress,showpacs]{revtex4-1}
\usepackage{bm}
\usepackage{amssymb}
\usepackage{colordvi}
\usepackage{graphicx}
\usepackage{color}
\usepackage{hyperref}
\usepackage{ulem}
\usepackage{color}

\newcommand{\beq}{\begin{equation}}
\newcommand{\eeq}{\end{equation}}
\newcommand{\be}{\begin{eqnarray}}
\newcommand{\ee}{\end{eqnarray}}

\def\ket#1{\vert #1 \rangle}
\def\bra#1{\langle #1 \vert}

\begin{document}

\title{Anomalous Quadrupole Topological Insulators in 2D Nonsymmorphic Sonic Crystals}

\author{Zhi-Kang Lin}
\affiliation{School of Physical Science and Technology, \&
Collaborative Innovation Center of Suzhou Nano Science and
  Technology, Soochow University, 1 Shizi Street, Suzhou 215006,
  China}
\author{Hai-Xiao Wang}
\affiliation{College of Physics and Technology, Guangxi Normal University, Guilin 541004, China}
\affiliation{School of Physical Science and Technology, \&
Collaborative Innovation Center of Suzhou Nano Science and
  Technology, Soochow University, 1 Shizi Street, Suzhou 215006,
  China}
\author{Zhan Xiong}
\affiliation{School of Physical Science and Technology, \&
Collaborative Innovation Center of Suzhou Nano Science and
  Technology, Soochow University, 1 Shizi Street, Suzhou 215006,
  China}
\author{Ming-Hui Lu}
\affiliation{National Laboratory of Solid State Microstructures and Department of Materials Science and Engineering, Nanjing University, Nanjing, 210093, China, \&
Collaborative Innovation Center of Advanced Microstructures, Nanjing University, Nanjing, 210093, China}
\author{Jian-Hua Jiang}\email{Corresponding author: jianhuajiang@suda.edu.cn}
\affiliation{School of Physical Science and Technology, \&
Collaborative Innovation Center of Suzhou Nano Science and
  Technology, Soochow University, 1 Shizi Street, Suzhou 215006,
  China}

\date{\today}

\begin{abstract}
The discovery of quadrupole topology opens a new horizon in the study of topological phenomena. However, the existing experimental realizations of quadrupole topological insulators in symmorphic lattices with $\pi$-fluxes often break the protective mirror symmetry. Here, we present a theory for anomalous quadrupole topological insulators in nonsymmorphic crystals without flux, using 2D sonic crystals with $p4gm$ and $p2gg$ symmetry groups as concrete examples. We reveal that the anomalous quadrupole topology is protected by two orthogonal glide symmetries in square or rectangular lattices. The distinctive features of the anomalous quadrupole topological insulators include: (i) minimal four bands below the topological band gap, (ii) nondegenerate, gapped Wannier bands and special Wannier sectors with gapped composite Wannier bands, (iii) quantized Wannier band polarizations in these Wannier sectors. Remarkably, the protective glide symmetries are well-preserved in the sonic-crystal realizations where higher-order topological transitions can be triggered by symmetry or geometry engineering.
\end{abstract}

\maketitle

Topological insulators are unconventional materials which host robust edge states and quantized transport or electromagnetic properties as dictated by the topological invariants of the occupied bulk bands~\cite{rev1,rev2}. These topological invariants result from nontrivial quantization of Berry's phases arising from parallel transport in the Brillouin zone, which are essentially connected to dipole polarization and topological charge pumping in crystals~\cite{pol1}. Though developed for electronic systems, topological band theory also applies for photonic~\cite{ph1,ph2,ph3,ph4,ph5,ph6,ph7,ph8,ph9,ph10,ph11,ph12,ph13} and mechanical~\cite{ac1,ac2,ac3,ac4,ac5,ac6,ac7,ac8,ac9,me1,me2} waves, yielding rich phenomena and applications in various disciplines of science.

\begin{figure}
\begin{center}
\includegraphics[width=3.4 in]{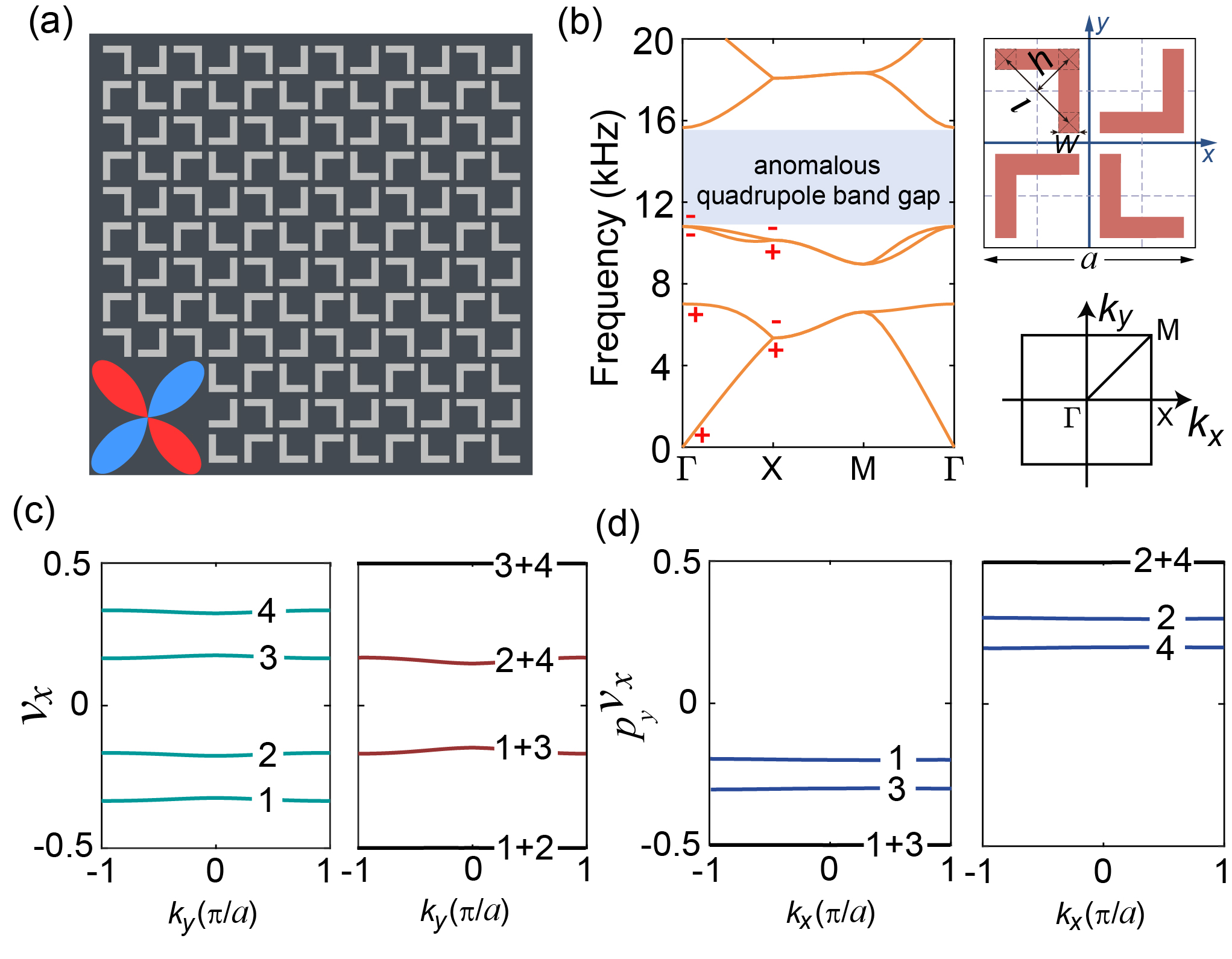}
\caption{(Color online)\ 
(a) A $p4gm$ SC that realizes the AQTI. Inset depicts the bulk quadrupole moment. (b) Acoustic band structure for $l=0.42a$, h=$0.21a$, $w=0.1a$, and $a=2$~cm. Symbols $\pm$ label parities at the $\Gamma$ and X points. Right panel: Unit-cell (brown shapes denote the epoxy scatterers, while dashed-lines depict the glide invariant lines) and Brillouin zone of the SC. (c) Wannier bands for the four acoustic bands below the gap (left), their combinations (right). (d) Nested Wannier bands and their combinations in the ``1+3'' and ``2+4'' Wannier sectors. Sound velocity and mass density of air are set as 343~m/s and 1.29 kg/m$^3$, respectively.
}
\end{center}
\end{figure}

Recently, topological band theory was generalized from conventional (dipole) topology to quadrupole and octopole band topology~\cite{qti1,qti2,exp1,exp2,exp3,exp4}, opening a pathway toward higher-order topology~\cite{hoti1,hoti2,hoti3,hoti4,hoti5,new1,new2,new3,new4,new5,new6,new7,aqti1}. For instance, a 2D quadrupole topological insulator (QTI) has gapped edge states and topologically-protected corner states. QTI was proposed theoretically using a $\pi$-flux tight-binding model with noncommutative mirror symmetries~\cite{qti1,qti2} and was later observed in mechanical~\cite{exp1} and microwave metamaterials~\cite{exp2}, $LC$-circuits~\cite{exp3}, and coupled optical waveguides~\cite{exp4}. However, the existing material realizations of QTIs often break the mirror symmetry that protects the quadrupole topology~\cite{exp1,exp2,exp3}.

In this work, we present a theory of anomalous QTI (AQTI) in nonsymmorphic crystals without gauge flux. Using 2D square-lattice sonic crystals (SCs) as material realizations, we find that the AQTI is a topological state protected by two orthogonal glide symmetries. Due to these glide symmetries, the AQTI phase has several features that are distinct from the QTI model proposed in Refs.~\cite{qti1,qti2}. These features include: (i) minimal four bands below the topological band gap which originates from the band-sticking effect and the parity inversions, (ii) nondegenerate, gapped Wannier bands and special Wannier sectors with gapped composite Wannier bands, (iii) quantized Wannier band polarizations in the special Wannier sectors. Using SCs, we demonstrate that higher-order topological transitions can be triggered by symmetry or geometry engineering. We further present a full symmetry analysis for square and rectangular wallpaper crystals in the pursuit of quadrupole topology in flux-threading-free systems. Our discovery can be directly generalized to other classical-wave systems, e.g., photonic crystals, mechanical metacrystals. An experimental demonstration of the $p4gm$ SC as the acoustic AQTI has been demonstrated by some of us in Ref.~\cite{ourexp} based on the theory in this work.

{\sl Nonsymmorphic $p4gm$ SCs}.---We start with the AQTI phase in the $p4gm$ SCs [see Fig. 1(a)]. In each unit-cell, four identical arch-shaped epoxy blocks, with width $w$, height $h$ and length $l$, serve as the scatterers for acoustic waves. The scatterers are arranged such that the SC has $p4gm$ nonsymmorphic symmetry group. The four key symmetries are: two orthogonal glide symmetries, $G_x=\{m_x|{\bf \tau}_y\}$ and $G_y=\{m_y|{\bf \tau}_x\}$ where $m_x:=x\to \frac{a}{2}-x$, $m_y:=y\to \frac{a}{2}-y$, ${\bf \tau}_y:=y\to y+\frac{a}{2}$ and ${\bf \tau}_x:=x\to x+\frac{a}{2}$ ($a$ is the lattice constant, and the origin of the coordinate is at the center of the unit-cell), the inversion symmetry ${\cal I}$ and the $C_4$ rotation symmetry.

The acoustic band structure, as obtained from COMSOL Multiphysics, is shown in Fig. 1(b). The band gap of concern is that between the fourth and the fifth bands. The dipole polarization of the bands below the gap is given by~\cite{qti2}, 
\be
P_x=P_y=\frac{1}{4}(1-\xi), \quad \xi=\prod_n {\cal I}_n(\Gamma){\cal I}_n(X),
\ee
where ${\cal I}_n(\Gamma)$ and ${\cal I}_n(X)$ are the parity eigenvalues of the $n^{th}$-band ($n=1,2,3,4$) at the $\Gamma$ and X points, respectively. From the parity eigenvalues in Fig.~1(b), the first and second (third and fourth) bands have a dipole polarization of ${\vec P}=(\frac{1}{2},\frac{1}{2})$. Therefore, the band gap of concern has a vanishing dipole polarization, which is necessary for the emergence of the quadrupole topology~\cite{qti1}.

The glide symmetries lead to the double degeneracy for all Bloch states on the XM and YM lines. This double degeneracy can be understood by constructing the anti-unitary operators $\Theta_j=G_j*{\cal T}$ ($j=x,y$) where ${\cal T}$ is the time-reversal operator (i.e., complex conjugation). It is straightforward to obtain that $\Theta_y^2\psi_{n,{\vec k}}=e^{ik_xa}\psi_{n,{\vec k}}$. Therefore, $\Theta_y^2\psi_{n,{\vec k}}=-\psi_{n,{\vec k}}$ at the XM line, i.e., $k_x=\frac{\pi}{a}$. As an analog to the Kramers theorem for fermions, such a relation results in the double degeneracy for all Bloch states on the XM line~\cite{ph10}. According to the $C_4$ symmetry, the YM line has the same property.

The parity operator ${\cal I}$ can be expressed using the glide operators, ${\cal I}={\bf \tau}_y^2G_x G_y ={\bf \tau}_x^2 G_y G_x$. Obviously, the two glide operators do not commute. Besides, we find that the double degeneracy at the $X$ point consists of Bloch states with opposite parities, since $\Theta_y{\cal I}=-{\cal I}\Theta_y$. This property constraints the parity-inversion between the $\Gamma$ and $X$ points and hence the dipole polarization. In this way, the quadrupole topology, since it requires a vanishing dipole polarization, emerge in our system when there are four bands below the band gap.

{\sl Wannier bands and nested Wannier bands.}---The quadrupole topology is characterized by two features: (i) gapped Wannier bands for both $\nu_x$ and $\nu_y$~\cite{qti1,qti2}; (ii) quantized Wannier band polarizations. The second feature requires the mirror or glide reflection symmetry with respect to both $x$ and $y$. The first condition, however, cannot be fulfilled in the presense of commutative mirror symmetries, as shown in Ref.~\cite{qti2}.

The Wannier bands are calculated from the Wilson-loop approach~\cite{hx}. For instance, $\exp[i2\pi\nu_x(k_y)]$, is obtained from the eigenvalues of the Wilson-loop operator $\hat{W}^x_{\vec k}={\cal T}_P\exp[i\oint \hat{A}^xdk_x]$ where the superscript $x$ and the subscript ${\vec k}$ denote the direction and starting point of the Wilson-loop, respectively. $\hat{A}^x$ is the non-Abelian (matrix) Berry connection for the first four bands. The matrix element is $A^x_{nm}=\bra{u_n({\vec k})}i\partial_{k_x}\ket{u_m({\vec k})}$ for $n,m=1,...,4$. Here, $u_n({\vec k})$ is the periodic part of the Bloch wavefunction, and ${\cal T}_P$ is the path ordering operator. The results in Fig.~1(c) show that there are four nondegenerate, gapped Wannier bands.

The symmetry constraints on the Wannier bands are (see Supplemental Material for detailed derivations),
\begin{subequations}
\begin{align}
& \nu_{x,n}(k_y) \overset{\underset{G_x}{}}{=} - \nu_{x,n^\prime} (k_y) + \frac{1}{2}\ {\rm mod}\ 1, \\
& \nu_{x,n}(k_y) \overset{\underset{G_y}{}}{=}  \nu_{x,n^\prime} (-k_y) + \frac{1}{2} \ {\rm mod}\ 1, \\
& \nu_{x,n}(k_y) \overset{\underset{{\cal I}}{}}{=} -\nu_{x,n^\prime}(-k_y)\ {\rm mod}\ 1, \\
& \nu_{x,n}(k_y) \overset{\underset{{\cal T}}{}}{=} \nu_{x,n}(-k_y)\ {\rm mod}\ 1.
\end{align}
\end{subequations}
Here, $n^\prime\ne n$ are Wannier band indices. We notice that ${\cal I}$ transforms the first (second) Wannier band to the fourth (third) Wannier band. Besides, $G_x$ transforms the first (third) Wannier band into the second (fourth) Wannier band, while $G_y$ transforms the first (second) Wannier band to the third (fourth) Wannier band. Thus, the four Wannier bands are intrinsically connected to each other.

Following Refs.~\cite{qti1,qti2}, for the description of the quadrupole topology, it is important to introduce a division of the Wannier bands below and above $\nu_y=0$. For the situations with more than two Wannier bands, such a division can have more possibilities. From the symmetry analysis above, we find that the division ``1+2" and ``3+4'' always yield gapless composite Wannier bands, i.e., $\nu_{x, 1}+\nu_{x, 2}=\nu_{x, 3}+\nu_{x, 4}=\frac{1}{2}~ {\rm mod}~ 1$ [see Fig.~1(c)]. Only the division ``1+3" and ``2+4'' leads to gapped composite Wannier bands (``gapped'' meaning that the Wannier bands are away from the special values of 0 and $\pm \frac{1}{2}$) [see Fig.~1(c)]. We denote the Wannier sectors, ``1+3" and ``2+4'', as I and II, respectively. The Wannier sector I (II) has a Wannier center below (above) the unit-cell center.

For Wannier sectors I and II, the Wannier band polarizations are calculated as $P^{\nu_{x},I}_{y}=\frac{1}{2\pi}\int dk_x p^{\nu_x}_{y,1}(k_x)+p^{\nu_x}_{y,3}(k_x)$ and $P^{\nu_{x},II}_{y}=\frac{1}{2\pi}\int dk_x p^{\nu_x}_{y,2}(k_x)+p^{\nu_x}_{y,4}(k_x)$, respectively. The nested Wannier bands, $p^{\nu_x}_{y,n}(k_x)$ ($n=1,2,3,4$), are calculated through the nested Wilson-loop approach (see Supplemental Material for details). The symmetry constraints on the Wannier band polarizations are (see Supplemental Material for proofs), 
\begin{subequations}
\begin{align}
& P^{\nu_x,I}_{y} \overset{\underset{G_x}{}}{=} P^{\nu_x,II}_{y} \ {\rm mod}\ 1, \\
& P^{\nu_x,I}_{y} \overset{\underset{G_y}{}}{=} - P^{\nu_x,I}_{y} \ {\rm mod}\ 1, 
\end{align}
\end{subequations}
The Wannier band polarizations is thus quantized as
\begin{align}
& P^{\nu_x,I}_{y}=P^{\nu_x,II}_{y} = 0, \frac{1}{2} \ {\rm mod}\ 1 .
\end{align}
Similarly, the polarizations are quantized for the Wannier bands $\nu_y(k_x)$, leading to $P^{\nu_y,I}_{x}=P^{\nu_y,II}_{x} = 0, \frac{1}{2} \ {\rm mod}\ 1$. These constraints are confirmed by numerical calculations. In the $p4gm$ SC, $P^{\nu_x,I}_{y}=P^{\nu_x,II}_{y}=\frac{1}{2} \ {\rm mod}\ 1$ [see Fig.~1(d)], leading to nontrivial quadrupole topological index
\be
q_{xy} = 2 P^{\nu_x,I}_{y} P^{\nu_y,I}_{x}=\frac{1}{2} \ {\rm mod}\ 1 .
\ee

\begin{figure}[htb]
\begin{center}
\includegraphics[width=3.45 in]{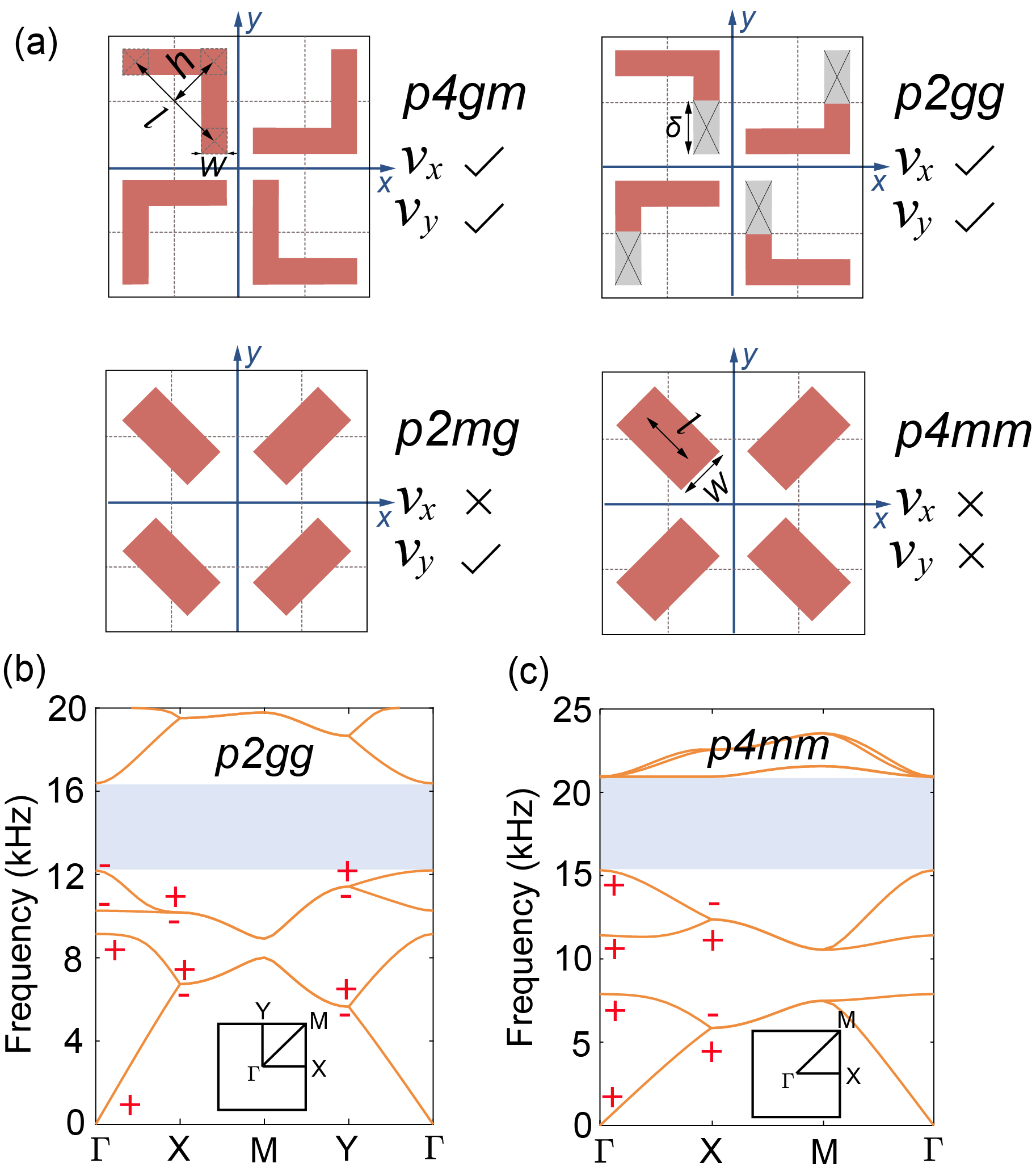}
\caption{(Color online)\ 
(a) Schematic of four types of SCs with $p4gm$, $p2gg$, $p2mg$ and $p4mm$ symmetries, separately. The brown shapes denote the epoxy scatterers, while the dashed-lines depict the glide invariant lines. We use the checkmarks to denote whether the Wannier bands $\nu_x$ and $\nu_y$ are gapped or not. (b)-(c) Acoustic band structures for the (b) $p2gg$ and (c) $p4mm$ SCs. The cyan regions denote the bulk band gaps of concern. Geometry parameters for (b): $l=0.5a$, $h=0.25a$, $w=0.05a$, and $\delta=0.1a$ ($a$ is the lattice constant), and for (c): $l=0.35a$, $h=0$, and $w=0.25a$. Here, the $p2gg$ SC is derived from the $p4gm$ SC by removing the gray regions of length $\delta$ for the epoxy scatterers [see figure (a)].}
\end{center}
\end{figure}

\begin{figure}
\begin{center}
\includegraphics[width=3.5 in]{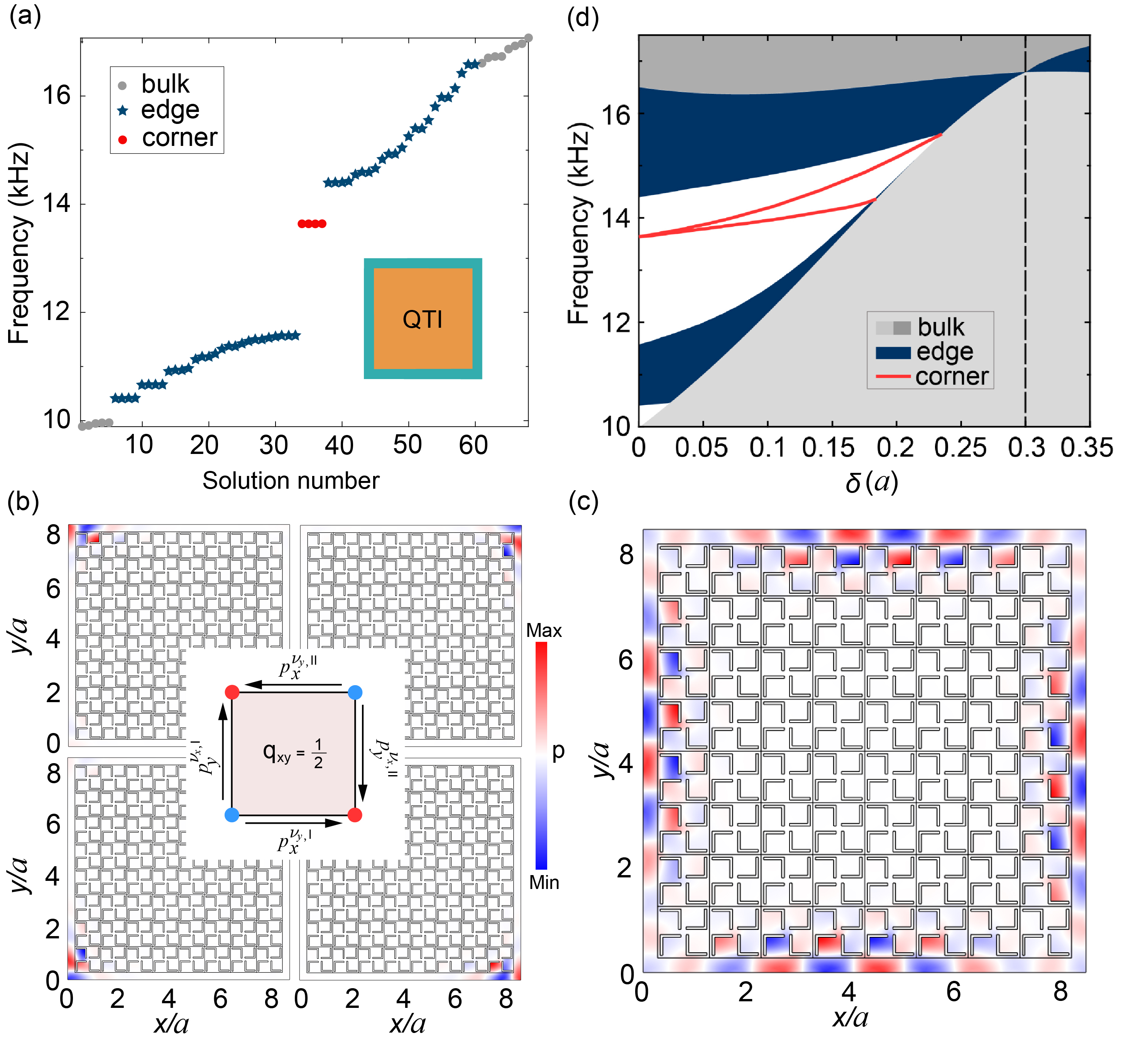}
\caption{(Color online)\ 
(a) Eigenstates spectrum for a square-shaped finite-sized $p4gm$ SC with hard-wall boundary conditions, showing the coexistence of the bulk, edge and corner states. (b)-(c): Acoustic wavefunction (the acoustic pressure field $p$) for the (b) corner and (c) edge states. (d) Evolution of the bulk, edge and corner spectra with the truncation length $\delta$ for the $p2gg$ SC. The topological band gap is tuned to close and reopen with increasing $\delta$. Bulk band gap closing at $\delta=0.3a$ is indicated by the dashed line.
}\end{center}
\end{figure}

{\sl Symmetry considerations}.---The above results, in particular, the quantization of the bulk quadrupole moment $q_{xy}$ is determined by the glide symmetries, $G_x$ and $G_y$. Therefore, these results are also applicable to $p2gg$ SCs ($p2gg$ is similar to $p4gm$ but without the $C_4$ rotation symmetry), as shown in next section. We now consider possible emergence of the AQTI phase in square/rectangular groups within the 17 wallpaper groups. Fig. 2(a) illustrates the structures of four types of SCs with $p4gm$, $p2gg$, $p2mg$ and $p4mm$ symmetries, separately. Analysis on other symmetry groups are presented in the Supplemental Material. We remark that although we often use square-lattices as examples, the analysis can be directly generalized to rectangular lattices. For the sake of continuity, the $p2gg$ SC is constructed from the $p4gm$ SC by truncating the gray regions as depicted in the figure. The $p2mg$ SC has a glide symmetry, $G_y$, and a mirror symmetry, $M_x:=(x,y)\to (-x,y)$. The $p4mm$ SC has two mirror symmetries, $M_x$ and $M_y:=(x,y)\to (x,-y)$. The unit-cells of the $p2mg$ and $p4mm$ SCs are doubled to enable comparison and analysis of the band structures and the Wannier bands for all SCs. The acoustic band structures for the $p2gg$ and $p4mm$ SCs are presented in Figs.~2(b) and 2(c), respectively. From the parity eigenvalues at the high-symmetry points of the Brillouin zone, we conclude that, for both SCs, the second band gap carries vanishing dipole polarization. However, the $p2gg$ SCs can have nontrivial quadrupole topology, whereas the $p4mm$ SCs cannot. The underlying reason is that the commutative mirror symmetries, $M_x$ and $M_y$, in the $p4mm$ SCs lead to degenerate Wannier bands and gapless composite Wannier bands. Similar results were found in the $p2mm$ and $c2mm$ wallpaper crystals. A special case is $p2mg$ crystals where the commutation between the mirror and glide symmetry at special high-symmetry points yields gapped Wannier bands for $\nu_y$, but gapless Wannier bands for $\nu_x$. These wallpaper crystals cannot support quadrupole topology. We find that to have nondegenerate, gapped Wannier bands, both glide symmetries $G_x$ and $G_y$ are needed, or if the system has neither glide nor mirror symmetry in the $x$ and/or $y$ direction. In the latter case (including $p2ll$, $p1ml$, $p1gl$, $c1ml$, $p4$ wallpaper crystals), however, the Wannier band polarizations are not quantized (at least for one of them, $\nu_x$ or $\nu_y$), because the quantization of the Wannier band polarizations are dictated by the glide or mirror symmetry. Therefore, only for the $p4gm$ and $p2gg$ SCs, the AQTI phase can emerge.

{\sl Edge and corner states.}---The nontrivial quadrupole topology leads to gapped edge states and in-gap corner states. To demonstrate these topological boundary states, we study a finite SC with $8\times 8$ unit-cells cladded by hard-wall boundaries. A finite-width ($0.28a$) air channel is introduced between the hard-wall boundaries and the SC to tune the edge band gap to the middle of the bulk band gap. Such a narrow air channel cannot introduce waveguide modes in the bulk band gap, since the waveguide modes can appear only at much higher frequencies ($>30$~kHz). In accordance with our theory, the calculated spectrum [see Fig.~3(a)] shows the emergence of four corner states in the common spectral gap of the edge and bulk. The acoustic wavefunctions of the corner and edge states are shown in Figs.~3(b) and 3(c), respectively. The four degenerate corner states exhibit the quadrupole symmetry.

\begin{figure}[htb]
\begin{center}
\includegraphics[width=3.45 in]{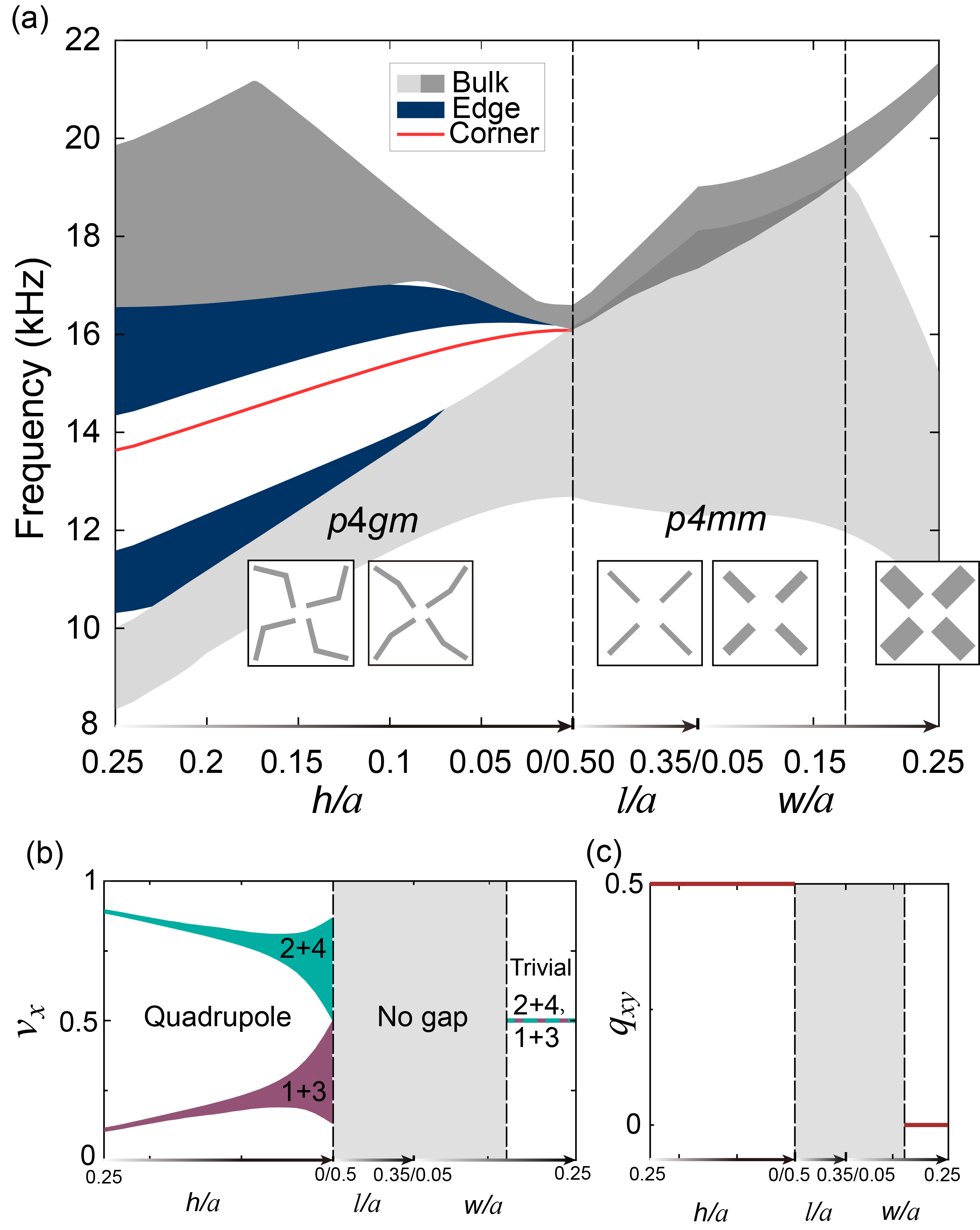}
\caption{(Color online)\ (a) Evolution of the bulk, edge and corner spectra during a geometry transformation from the $p4gm$ group to the $p4mm$ group: The starting point has $l=0.5a$, h=$0.25a$, and $w=0.05a$. First, $h$ goes from $0.25a$ to 0. Then, $l$ goes from $0.5a$ to $0.35a$. Finally, $w$ goes from $0.05a$ to $0.25a$. The vertical dashed lines indicate the conditions where the bulk band gap closes or opens. The left, middle and right regions separated by the dashed lines correspond to the QTI phase, the gapless phase, and the trivial phase, separately. Several representative unit-cell geometries are illustrated as the insets. (b)-(c) Evolution of the Wannier bands (b) and the quadrupole topological
index $q_{xy}$ (c) during the geometry transformation process.
}
\end{center}
\end{figure}

{\sl Geometry engineering in the $p2gg$ group}.---Fig.~3(d) presents the evolution of the bulk, edge and corner states as a function of the geometry parameter $\delta$. For $\delta=0$, the SC has $C_4$ symmetry and belongs to the $p4gm$ group where all four corner states are degenerate. As $\delta$ increases, the degeneracy is lifted and split into two doublets: the left-upper and right-lower corner states form a doublet, while the right-upper and left-lower corner states form another doublet. At large $\delta$, the edge band gap is closed ($\delta=0.233a$), while the corner states merge into the edge, i.e., their wavefunctions gradually transform from fully localized to extensive along the edges. The closure of the edge band gap signals the transition into a quadrupole trivial phase. Further increase of $\delta$ leads to bulk band gap closing at $\delta=0.3a$ where the edge states merge into the bulk (see Supplemental Material for details).

{\sl Symmetry engineering}.---Starting from the $p4gm$ SC, one can keep the $C_4$ symmetry and gradually transform into the $p4mm$ SC, by reducing the geometry parameter $h$. We study a continuous transformation process with first (i) $h$ going from $0.25a$ to 0, and then (ii) $l$ going from $0.5a$ to $0.35a$, and lastly (iii) $w$ going from $0.05a$ to $0.25a$. The phase diagram for the corner, edge and bulk spectra are shown in Fig.~4(a) which indicates a topological transition taking place exactly at the geometry transition point, i.e., $h=0$. While the SC is transformed from the $p4gm$ group to the $p4mm$ group, the bulk band gap closes and the corner states merge into the bulk. The edge states remain gapped and gradually merge into the bulk bands before the bulk band gap closing. The bulk band gap remains closed in stage (ii). When the band gap is reopened in stage (iii), the corner states disappear in the bulk band gap. The phase diagram in Fig.~4(a) thus manifests directly the bulk-corner correspondence due to the quadrupole topology.

Such a topological transition is also manifested in the Wannier bands. As shown in Fig.~4(b), the Wannier band gap for the ``1+3'' and ``2+4'' sectors is closed when the SC expriences the symmetry transition from the $p4gm$ group to the $p4mm$ group. We find that the $p4gm$ SCs have nontrivial quadrupole topology, i.e., $q_{xy}=\frac{1}{2}$ (see Fig.~4(c); This is confirmed in a recent experiment~\cite{ourexp}). In contrast, when the band gap is reopened in the $p4mm$ SC, the Wannier bands are degenerate which cannot support the quadrupole topology.

Finally, we remark that the acoustic topological band gaps studied here do not have the chiral symmetry. Consequently, the corner states are not fixed at the mid-gap frequency. However, as shown in Ref.~\cite{bic}, this property can be used to move the corner states into the bulk continuum where the corner states still survive.

{\sl Note added}: At the final stage of this work, we became aware of recent works on boundary obstructed topological phases~\cite{obstructed} (including some quadrupole topological insulators) and quadrupole topological insulators in magnetized systems~\cite{mQTI1,mQTI2}.

{\sl Acknowledgments.}---
J.H.J, Z.K.L, H.X.W, and Z.X acknowledge supports from the Jiangsu specially-appointed professor funding and the National Natural Science Foundation of China (Grant No. 11675116). M.H.L thanks the support from the National Key R\&D Program of China (2017YFA0303702 and 2018YFA0306200). J.H.J thanks Prof. Gil Young Cho for illuminating discussions.

\end{document}